\documentstyle[prl,aps,twocolumn]{revtex} 
\begin{document} \twocolumn[\hsize\textwidth\columnwidth\hsize\csname @twocolumnfalse\endcsname\draft
\title{Correlation effects in the ground state charge density of Mott-insulating NiO:
a comparison of {\it ab-initio} calculations and high-energy electron diffraction 
measurements} 
\author{S. L. Dudarev$^{1,\dagger}$, L.-M.  Peng$^{2,\ddagger}$, 
S.  Y.  Savrasov$^1$ and J.-M.  Zuo$^3$}
\address{$^1$Max-Planck-Institut f\"ur Festk\"orperforschung, Heisenbergstra\ss e 1, D-70569
Stuttgart, Germany} \address{$^2$Department of Electronics, Peking University, Beijing 100871, China}
\address{$^3$ Department of Physics and Astronomy, Arizona State University, Tempe, AZ 85287, USA}
\date{\today} \maketitle

\begin{abstract} Accurate high-energy electron diffraction measurements of structure factors of NiO
have been carried out to investigate how strong correlations in the Ni 3{\it d} shell affect electron
charge density in the {\it interior} area of nickel ions and whether the new {\it ab-initio}
approaches to the electronic structure of strongly correlated metal oxides are in accord with
experimental observations.  The generalized gradient approximation (GGA) and the local spin density
approximation corrected by the Hubbard U term (LSDA+U) are found to provide the closest match to
experimental measurements.  The comparison of calculated and observed electron charge densities shows
that correlations in the Ni 3{\it d} shell suppress covalent bonding between the oxygen and nickel
sublattices.  
\end{abstract} 
\date{\today} 
\vskip 0.5cm ]

Recent years have witnessed the largely unexpected \cite{Christian1975} progress in the development of
computational approaches to the evaluation of fundamental properties of materials from the first
principles.  The stimulus for this development was provided by the Hohenberg-Kohn theorem
\cite{Hohenberg1964} which establishes that the energy of the ground state of a solid is a functional
of its one-electron density $\rho ({\bf r})$.  The problem of accurate determination of $\rho ({\bf
r})$ therefore acquires fundamental significance for the physics of materials {\it both} from the
experimental {\it and} theoretical points of view.  In cases where accurate experimentally measured
and calculated charge densities are available (like, e.g.  in the case of silicon \cite{JMZuo97}), the
quality of {\it ab-initio} approximations can be assessed on the basis of the agreement between
experimental and theoretical data.

The Kohn-Sham method \cite{Kohn1965}, which provides a convenient way of carrying out density
functional calculations, in certain cases encounters serious difficulties.  For example, it predicts
metallic ground states for a number of late transition metal monoxides where metal ions have partly
filled electronic shells.  Nickel and cobalt monoxides are often quoted\cite{Fulde1995} as typical
examples illustrating the failure of conventional density functional methods to describe the effective
one-particle band structure of Mott insulating materials\cite{Terakura1984,Kubler1986}.  Several
modified density functional schemes have been proposed lately to explain the nature of large bandgaps
observed for CoO and NiO.  These schemes include the orbital polarization correction
\cite{Norman1990}, the self-interaction correction (SIC) \cite{SIC} and the local spin density
approximation including the Hubbard U term (LSDA+U) \cite{Anisimov1991,Liechtenstein1995}.  The new
approximations improve the description of the effective one-particle band structure of Mott
insulators, and their validity is further confirmed by the recent studies of orbital ordering in
transition metal compounds, see e.g.  \cite{Ezhov1998}.

At the same time it is widely appreciated that the new `improved' computational schemes represent a
departure from the original formulation of density functional theory \cite{Hohenberg1964}.  The new
approximations employ functionals that depend not only on the spin density of electrons $\rho _{\sigma
}({\bf r})$ but also on the {\it orbital} occupation numbers that in turn depend on the choice of the
basis functions.  Calculations performed using the new {\it ab-initio} schemes result in better values
for bandgaps and magnetic moments \cite{Anisimov1991}.  At the same time the use of the modified
energy functionals alters the relative occupancies of {\it d}-states \cite{WEPick98} and alters the
distribution of the charge density in the unit cell.  This raises the question of how well the new
functionals describe the main entity of density functional theory, namely, $\rho ({\bf r})$
itself.

In this paper we investigate this issue by comparing the calculated and experimentally observed charge
density distributions.  We compare the structure factors of NiO that were measured using a recently
developed highly accurate electron diffraction technique \cite{Zuo1988,review,Zuo1997} and calculated
theoretically using several {\it ab-initio} linear muffin-tin orbital (LMTO)-based methods
\cite{Andersen1975,Savrasov1992}, including the LSDA+U approach.

At present, there is no sufficiently accurate experimental information on the distribution of electron
charge density in the unit cell of NiO or other similar transition metal oxides.  The powder X-ray
diffraction techniques that were successfully used to determine the equilibrium positions of ions in a
unit cell \cite{Rooksby1948,Massarotti1991}, do not have the accuracy required for observing the
relatively small changes in the charge density resulting from the competition between covalent bonding
and correlation effects.  The convergent beam electron diffraction (CBED) technique used here takes
advantage of the fact that electron beam can be focused on a small nearly perfect area of the sample
and the resulting diffraction pattern can be simulated using highly accurate multiple scattering
dynamical diffraction approach \cite{review} therefore eliminating the extinction problem that limits
the accuracy of X-ray techniques.  The high precision of electron diffraction measurements has made it
possible to study subtle details of the charge density distribution in band insulators like MgO
\cite{Zuo1997} and Cu$_2$O \cite{JMZuo99}.  In this paper we for the first time investigate how the
interplay between covalent bonding and the Coulomb on-site repulsion between {\it d}-electrons in a
partially filled shell influences the charge density distribution in a crystal unit cell of a Mott
insulating oxide.

Nickel monoxide is probably the most extensively studied Mott insulating material
\cite{Brandow1977,Hufner1994} and it is often referred to as the prototype of the entire class of
`anomalous' transition metal oxides.  Depending on the type of approximation used in an {\it
ab-initio} calculation, the ground state of NiO is predicted to be a metal (LDA, non-magnetic state),
a $\sim$0.4 eV bandgap Mott insulator (LSDA, antiferromagnetic state), a $\sim$1.0 eV bandgap Mott
insulator (GGA, antiferromagnetic state) or a $\sim$3.0 eV bandgap charge-transfer insulator (LSDA+U,
antiferromagnetic state).

\vskip 2cm \centerline{Figure 1 near here} \vskip 2cm 

Fig.1 shows plots of the density of states calculated for the above four cases.  The X-ray
photoelectron spectroscopy data \cite{Hufner1994} agree best with the LSDA+U one-particle band
structure \cite{Anisimov1991,Anisimov1993} that shows that NiO is a charge-transfer insulator where
the bandgap separates filled oxygen 2{\it p} and empty nickel 3{\it d} states.  The band structures of
NiO calculated using either LSDA or GGA show instead that the bandgap separates filled and empty
nickel 3{\it d} states and that NiO is therefore a Mott-Hubbard insulator.

To characterize the distribution of electron charge density in a unit cell of NiO, we measured seven
low order energy dependent structure factors \cite{Hirsch1977} $U({\bf G})$ that are defined by
\begin{eqnarray} U({\bf G})&=&{2me^2 \over \pi \hbar ^2 G^2 \Omega } \left (\int \limits _{\Omega}\rho
^{(T)}({\bf r}) \exp (-i {\bf G}\cdot {\bf r})d^3r \right.  \nonumber \\ &-&\left .\sum _{\alpha }
Z_{\alpha } \exp (-i {\bf G}\cdot {\bf r}_{\alpha }) \exp \left [-{1\over 2} \langle ({\bf G}\cdot
{\bf u}_{\alpha })^2\rangle\right ]\right) \label{1} \end{eqnarray} where ${\bf G}$ is a reciprocal
lattice vector, $m$ is the relativistic electron mass and $\Omega $ is the volume of the unit cell.
Summation over $\alpha $ is performed over ions in a unit cell, $Z_{\alpha}$ is the charge, ${\bf
r}_{\alpha}$ is the equilibrium position and ${\bf u}_{\alpha}$ is the thermal displacement of the
respective nucleus.  $\rho ^{(T)}({\bf r})$ is the electron density averaged over the thermal
ensemble.

The experiment was performed using the LEO-912 $\Omega $ energy-filtering electron microscope with the
Gatan liquid nitrogen cooled sample holder.  The specimen used is a single crystal NiO cooled to about
110K.  The small rhombohedral distortion of NiO at 110 K was measured using higher-order Laue zone
lines \cite{Eaglesham1988,JMZuo98} to be $a = 4.25 $\AA \ and 
$\alpha = 90.044 ^{\circ}$.  This small
distortion was neglected in the charge density study.  The experimental CBED patterns were recorded
using a 15 eV energy-filtering slit that was placed around the zero-loss peak.  This was done to
remove the contribution from inelastically scattered electrons that form continuous background due to
plasmon and higher energy loss processes.  Off-zone-axis systematic diffraction conditions were used
to collect diffraction intensities for reflections up to (440).  The experimental patterns recorded
using a slow-scan charge coupled device (CCD) camera were processed for the subsequent fitting using
procedures described in \cite{review}.  The refined values of the structure factors were obtained
using the EXTAL program \cite{review}.  Fig.2 illustrates the level of agreement between the measured
intensity variations and multiple scattering dynamical diffraction simulations used in the refinement
procedure.  The error in the measured structure factors was estimated by comparing the results
obtained using line scans taken at different positions (see Fig.2).

The calculated values of structure factors (\ref{1}) were obtained by integrating the self-consistent
solutions $\rho ({\bf r})$ of the Kohn-Sham equations found using the LMTO method \cite{Andersen1975}
and various approximations for the exchange-correlation potential.  Calculations were performed using
three $\kappa-$panels and 512 k-points in the Brillouin zone.  Convergence of the calculated values of
structure factors was ensured by varying the number of k-points and by introducing additional
approximations (e.g.  by taking into account the spin-orbit coupling).  The exchange-correlation
functionals were taken from \cite{LSDA} (LSDA), \cite{GGA96} (GGA) and \cite{LSDAU} (LSDA+U).  The
full self-consistent charge density was represented by a sum of two terms\cite{Savrasov1992} where the
first term $\tilde \rho ({\bf r})$ approximates the density in the region between the muffin-tin
spheres and is continuous across the boundaries of the spheres.  The second term approximates the
density inside the muffin-tine spheres and is represented by a spherical harmonics expansion
$Y_{lm}(\theta , \phi )$.  Substituting this in (\ref{1}) we arrive at \begin{eqnarray} &&\int \limits
_{\Omega}\rho ^{(T)}({\bf r}) \exp (-i {\bf G}\cdot {\bf r})d^3r \nonumber \\ &=&\exp \left [-{1\over
2} \langle ({\bf G}\cdot {\bf u})^2\rangle\right ] \int \limits _{\rm int.}\tilde \rho ({\bf r}) \exp
(-i {\bf G}\cdot {\bf r})d^3r \nonumber \\ &+&4\pi \sum _{\alpha }\exp (-i {\bf G}\cdot {\bf
r}_{\alpha }) \exp \left [-{1\over 2} \langle ({\bf G}\cdot {\bf u}_{\alpha })^2\rangle\right ]
\label{2} \nonumber \\ &\times &\sum _{l,m} Y_{lm}(\theta _{\bf G}, \phi _{\bf G}) \int \rho
^{(\alpha)}_{lm}(r) j_l(Gr) r^2 dr, \end{eqnarray} where $j_l(Gr)$ is a Bessel function, and ${\bf u}$
denotes the effective amplitude of thermal vibrations characterizing the motion of electrons in the
interstitial region.  No other thermal effects are taken into account in (\ref{2}) in accord with
spectroscopical data\cite{Tjernberg1996} showing no detectable changes in the electronic structure of
NiO occurring in the temperature range between 0$^{\circ}$K and 615$^{\circ}$K.

A meaningful comparison between experiment and theory has to take into account the thermal vibrations
of atomic nuclei.  For example, the value of the (111) structure factor calculated in the LSDA(AF)
approximation assuming that nuclei are frozen in their equilibrium positions equals $-4.488\cdot
10^{-2}$ \AA$^2$ to be compared with $-4.597\cdot 10^{-2}$ \AA$^2$ that was obtained assuming that
$B_{Ni}=0.131$ \AA$^2$ and $B_{O}=0.239$ \AA$^2$.  The difference between the two values is
approximately ten times the experimental uncertainty in the determination of this structure factor
(see Table) and this illustrates the significance of taking thermal vibrations into account.  However,
no reliable independent X-ray or neutron diffraction measurements on the Debye-Waller factors of Ni
and O ions is available in the literature (notably, {\it negative} values of $B_{Ni}$ were reported in
a recent publication \cite{Massarotti1991}).  To provide a starting approximation for a subsequent
refined search, we calculated the temperature factors of Ni and O using the shell model.

The shell model used for the theoretical evaluation of the Debye-Waller factors was developed by the
Chalk River group \cite{AWoods60}.  In this model an ion is represented by a massive core and a rigid
shell describing valence electrons.  Nine parameters in total were introduced to describe the lattice
dynamics of NiO, these include the shell charges, force constants for springs connecting the cores and
the shells as well as the first nearest neighbours (for the Ni$^{2+}$ ions) and up to the second
nearest neighbours (for the O$^{2-}$ ions).  The model parameters were obtained by fitting the
calculated phonon dispersion curves to the experimentally measured ones \cite{HBilz79}.  For the
chosen set of parameters of the model, we calculated the average thermal displacements $\langle {\bf
u}^2 \rangle$ for all the modes of lattice vibrations, and also the Debye-Waller factors for both the
nickel and oxygen ions.  Fig.3 shows the fitted phonon dispersion curves plotted for the $\langle
111\rangle$ and $\langle 110\rangle $ directions, and also the temperature dependent Debye-Waller
factors.  We have also investigated several other implementations of the shell model but found that
they led to no significant improvement in the description of the phonon dispersion curves
\cite{HGao99}.

To find more accurate values of the Debye-Waller parameters for each choice of the
exchange-correlation potential used for {\it ab-initio} calculations we plotted two-dimensional maps
of the reliability factor $R$ treating $B_{Ni}$ and $B_O$ as independent variables.

\vskip 2cm \centerline{Figure 4 near here} \vskip 2cm

The values of $B_{Ni}$ and $B_O$ corresponding to the minimum of the $R$-factor were then used to
obtain the values of $U({\bf G})$ shown in the Table.  Results listed in the Table show that the
estimated values of $B_{Ni}$ and $B_O$ are nearly independent on the choice of approximation used in
{\it ab-initio} calculations and that the spread of values of the Debye-Waller factor does not exceed
6\%.  The value of the Debye-Waller factor characterizing the thermal motion of electrons in the
interstitial region was evaluated using two different approximations, namely, $\langle
B\rangle=(B_O+B_{Ni})/2$ or $\langle B\rangle=(M_OB_O+M_{Ni}B_{Ni})/(M_O+M_{Ni})$. 
The difference
between structure factors evaluated using these two approximations was 
found to be significantly
smaller than the uncertainty of experimentally measured values 
of structure factors.

Apart from values calculated using the superposition of 
atomic densities, all the {\it ab-initio}  methods exhibit  
high (better than 1\%) degree of accord with experimental
data, with the exception of (111) structure factor. There is
a large spread among theoretical values of the (111) structure factor, 
which increases significantly with the inclusion of U. The structure
factor of (111) is most sensitive to the changes in the
distribution of the density of valence electrons, and the
differences among the theoretical models shows primarily the 
differences in the calculated ground state valence charge density.
In terms of the overall R-factor, the closest approximation to 
the experiment is provided by the generalized gradient approximation 
and by the LSDA+U approach (see Table). Better agreement with the
experimental value of the (111) structure factor is achieved with
the GGA.

The difference between the GGA and the LSDA+U charge density distributions is illustrated in Fig.5
where we mapped the densities calculated using the GGA and the LSDA+U approximations, subtracting from
each of them the density corresponding to the non-magnetic LDA solution.  \vskip 2cm
\centerline{Figure 5 near here} \vskip 2cm Fig.5 shows that the symmetry of the deformation of
electron density resulting from correlation effects remains the same both in LSDA+U and in GGA.  At
the same time there are significant differences in the radial structure of the density distributions
around Ni ions calculated for the two cases.  The LSDA+U approximation treats the wave functions of
3{\it d} states as `rigid' objects where the Hubbard correction shifts the filled and unoccupied
states in the opposite directions along the energy axis\cite{Anisimov1993,LSDAU}.  In the GGA
approximation the shape of wave functions and the population of the 3{\it d} states depends on the
local density and its gradient in the interior area of nickel ions.  The LSDA+U approximation relies
to a larger extent on the model assumptions and on the choice of tight-binding orbitals used for
treating electron correlations in a partly filled 3{\it d} shell.  The GGA approximation uses the
one-electron orbitals as auxiliary entities required in a calculation of total density $\rho ({\bf
r})$, which is the quantity observed experimentally using high-energy electron diffraction.

Fig.5 also shows a low resolution difference map between the experimentally observed and the GGA charge density
distribution estimated using seventy six low-order experimentally measured structure factors listed in
the Table (this includes transpositions and mirror reflections).  The comparison between the
experimental and calculated distributions confirms the trends revealed by the GGA and LSDA+U analysis
showing that correlation effects are responsible for the suppression of covalent bonding between the
metal and oxygen sublattices (this effect manifests itself in the reduction of the charge density in
the areas between the oxygen and nickel ions).  This agrees with the analysis of a similar effect
discovered in \cite{UO2} for uranium dioxide.  The charge density in a unit cell of {\it real} NiO is
more concentrated around atomic nuclei and it also shows tendency towards increasing in the region
between ions of the same type.  Qualitatively this may be interpreted as an indication that Ni {\it
d}-orbitals in fact have the shape that is different from that predicted by either the GGA or the
LSDA+U calculations.

There are several reasons responsible for the observed inaccuracy 
of {\it ab-initio} approaches.  For example, the LSDA+U approximation 
is based on the mean-field treatment of correlation
effects\cite{Anisimov1991}.  There could be other, 
more fundamental, circumstances leading to the
difference between the calculated and experimentally observed 
charge density distributions.  One of
the ideas behind the development of more accurate approaches 
to the treatment of electron correlations
in transition metal compounds consists in that the new approaches 
are intended to be used for
evaluating the parameters entering tight-binding {\it many-body} 
models of electron-electron
interactions.  These models are always based on a particular choice of orbitals associated with each
of the ions in the solid.  Our results show that the accuracy of the assumption that the charge
density may be decomposed into contributions associated with individual ions, is limited, and this
conclusion agrees with the analysis performed in Ref.\cite{WEPick98}.  To describe correlation effects
in oxides where covalency as well as correlation effects play a significant part, it may be necessary
to take into account changes in {\it both} the shape {\it and} occupation of localized electronic
orbitals.  A second-quantized many-body model describing intersite hopping and on-site Coulomb
interaction between electrons may prove to be sufficient for accounting for the positions of the main
peaks in the spectrum of excited states of an oxide.  At the same time even the exact solution of the
model may not be capable of giving a sufficiently accurate description to the the ground-state
properties of the system such as the distribution of the charge density of electrons in a unit cell.
The approach developed in this paper can also now be applied to test the accuracy and to compare
several other {\it ab-initio} methods that we did not consider above, for example, the
self-interaction correction\cite{SIC} or the first-principles Hartree-Fock approximation
\cite{Towler1994}.

In summary, by combining a recently developed high accuracy electron diffraction technique with {\it
ab-initio} calculations, we investigated how electron correlations in the Ni 3{\it d} shell influence
the distribution of charge density in the unit cell of NiO.  By comparing the experimentally measured
values of structure factors with values calculated using several different {\it ab-initio} approaches
we found that the structure factors evaluated using the generalized gradient approximation and the
LSDA+U approach agree best with the available experimental information.  The experimental data show
that the degree of covalent bonding in NiO is smaller than that predicted by theoretical calculations.

The authors would like to thank O. K. Andersen, A. P. Sutton and S. D. Kenny for stimulating 
discussions. SLD would like to acknowledge financial support from Deutscher Akademischer 
Austauschdienst during his visit to Max-Planck-Institut f\"ur Festk\"orperforschung, Stuttgart.  
The experimental measurements carried out at ASU HREM center were supported by NSF grant DMR 9412146.  
Calculations were performed in the Materials Modelling Laboratory of the Department of Materials 
at the University of Oxford, UK, using an HP Exemplar V-class computer jointly funded by 
Hewlett-Packard and HEFCE through the JREI scheme.  The work was also supported by the EPSRC JREI 
grant GR/M34454 and the National Science Fundation of China.

\vskip  2mm  $^{\dagger}$Permanent  address:  Department  of
Materials, University of Oxford, Parks Road, Oxford OX1 3PH,
UK

$^{\ddagger}$Also   at:  Beijing   Laboratory   of  Electron
Microscopy,   Center  for  Condensed   Matter   Physics  and
Institute of Physics, Chinese Academy of Sciences, P.O.  Box
2724, Beijing 100080, China

\section*{Figure captions}

\begin{figure} \caption{The total density of states plots of
NiO calculated  assuming (a) a non-magnetic ground state and
the local  density  approximation  (b) the type II AF ground
state and the local spin density  approximation (c) the type
II  AF   ground   state   and   the   generalized   gradient
approximation  (d) the type II AF ground state and the local
spin density  approximation  corrected by the Hubbard U term
(LSDA+U).  $\epsilon   _F$  denotes  the  Fermi  energy  and
$\epsilon  _C  ^{exp}$  shows  the  experimentally  observed
position   of   the   bottom   of  the   conduction   band.}
\label{Fig1} \end{figure}

\begin{figure}  \caption{An example of NiO structural factor
measurement  using  convergent  beam  electron   diffraction
(CBED).  The top is the experimentally  recorded diffraction
pattern  with  (220)  and  (440)  strongly  diffracted.  The
structural  factors  of (220) and  (440)  were  obtained  by
fitting intensities along the indicated lines.  The best fit
is  shown  in  bottom  right.  The   schematic   diagram  on
bottom-left  shows the  formation  of CBED by  focusing  the
electron   beam  on  the  top  of   crystalline   specimen.}
\label{Fig2} \end{figure}

\begin{figure} \caption{Plots illustrating the fitting of the
phonon   dispersion   curves   along   $\langle 111\rangle $  
and  $\langle 110\rangle $ directions, and the comparison  
between the values of the Debye-Waller factors evaluated using
the shell model and found by comparing the experimentally
measured and calculated structure factors.}
\label{Fig3} \end{figure}

\begin{figure}  \caption{A map showing the dependence of the
R-factor on the two Debye-Waller  factors $B_{Ni}$ and $B_O$
characterizing the amplitude of thermal vibrations of Ni and
O  ions   in   NiO.  Values   $B_{Ni}=0.133$   \AA$^2$   and
$B_{Ni}=0.244$   \AA$^2$   correspond   to  the  minimum  of
$R=0.0049$.}  \label{Fig4} \end{figure}

\begin{figure}  \caption{Cross-sections  of  charge  density
distribution in the (100) plane of NiO calculated  using the
GGA and LSDA+U approximations.  The two contour maps on the 
left-hand side show the difference between the self-consistent 
density distributions and the density calculated using the 
local density approximation for a non-magnetic ground state. 
The map on the right-hand side shows the difference between 
the experimentally observed charge density distribution and 
the distribution calculated using the generalized gradient 
approximation.} 
\label{Fig5} \end{figure}

\onecolumn

Table.  The  observed  and  calculated  values of  structure
factors  for  NiO.  The  Debye-Waller   temperature  factors
$B_{Ni}=8\pi   ^2\langle   u^2_{Ni}\rangle$   and  $B_O=8\pi
^2\langle  u^2_{O}\rangle$  for  each  calculated  set  were
introduced  following the  procedure  described in the text.
The energy of the incident electrons equals $E_0=119.5$ keV.
All the  values  listed  in the  table  are  given  with the
opposite    sign    and    in   \AA    $^2$    units    (see
Ref.\cite{review}).  The   $R$-factor   is   defined  as
$R=\sum  _{\bf  G}  W_{\bf  G}|U^{\rm  th}({\bf   G})-U^{\rm
exp}({\bf  G})|/  |U^{\rm  exp}({\bf  G})|$,  where  $U^{\rm
th}({\bf G})$ are the calculated and $U^{\rm  exp}({\bf G})$
are the experimentally  measured values.  The weight factors
$W_{\bf  G}$  are  given  by  $W_{\bf   G}=\sigma  _{\bf  G}
^{-1}/(\sum  _{\bf G} \sigma _{\bf G} ^{-1})$  where $\sigma
_{\bf G}$ represent  experimental  uncertainties. The error-bar
of the $R$-factor, $\delta R$, is given by
$(\delta R)^2=\sum  _{\bf  G}  W_{\bf  G}\sigma ^2 _{\bf G}
/  |U^{\rm  exp}({\bf  G})|^2$. Abbreviations  NM and AF
refer  to   non-magnetic   and   antiferromagnetic   states,
respectively.    \vskip    0.5cm    \begin{center}    \small
\vspace{0.5cm}   \begin{tabular}{|c|c|c|c|c|c|c|c|}   \hline
\mbox{ h k l} & observed  values (std.  dev.  $\sigma  _{\bf
G}$) & ATOMS & LSDA  (NM) & LSDA  (AF) &  LSDA+U  (AF) & GGA
(AF) & GGA+U (AF) \\ \hline B$_{Ni}$(\AA $^2$) & 0.135$^*$ &
0.131 & 0.129 & 0.131 & 0.136  &  0.133  & 0.137  \\  \hline
B$_{O}$  (\AA  $^2$) &  0.238$^*$  & 0.237 & 0.235 & 0.239 &
0.247   &   0.244   &   0.251   \\   \hline    $-U(111)$   &
4.632$\cdot10^{-2}$              ($\pm$0.012$\cdot10^{-2}$)&
4.401$\cdot10^{-       2}$      &       4.555$\cdot10^{-2}$&
4.597$\cdot10^{-2}$&                    4.668$\cdot10^{-2}$&
4.642$\cdot10^{-2}$ & 4.708$\cdot10^{-2}$\\ \hline $-U(200)$
&      9.083$\cdot10^{-2}$       ($\pm$0.022$\cdot10^{-2}$)&
9.486$\cdot10^{-       2}$      &       9.187$\cdot10^{-2}$&
9.204$\cdot10^{-2}$&                    9.181$\cdot10^{-2}$&
9.173$\cdot10^{-2}$ & 9.151$\cdot10^{-2}$\\ \hline $-U(220)$
&      6.640$\cdot10^{-2}$       ($\pm$0.036$\cdot10^{-2}$)&
6.756$\cdot10^{-       2}$      &       6.709$\cdot10^{-2}$&
6.716$\cdot10^{-2}$&                    6.705$\cdot10^{-2}$&
6.703$\cdot10^{-2}$ & 6.694$\cdot10^{-2}$\\ \hline $-U(311)$
&      2.482$\cdot10^{-2}$       ($\pm$0.010$\cdot10^{-2}$)&
2.482$\cdot10^{-       2}$      &       2.482$\cdot10^{-2}$&
2.482$\cdot10^{-2}$&                    2.481$\cdot10^{-2}$&
2.481$\cdot10^{-2}$ & 2.481$\cdot10^{-2}$\\ \hline $-U(222)$
&      5.187$\cdot10^{-2}$       ($\pm$0.026$\cdot10^{-2}$)&
5.336$\cdot10^{-       2}$      &       5.325$\cdot10^{-2}$&
5.318$\cdot10^{-2}$&                    5.307$\cdot10^{-2}$&
5.311$\cdot10^{-2}$ & 5.302$\cdot10^{-2}$\\ \hline $-U(400)$
&      4.456$\cdot10^{-2}$       ($\pm$0.018$\cdot10^{-2}$)&
4.427$\cdot10^{-       2}$      &       4.455$\cdot10^{-2}$&
4.456$\cdot10^{-2}$&                    4.457$\cdot10^{-2}$&
4.456$\cdot10^{-2}$ & 4.457$\cdot10^{-2}$\\ \hline $-U(440)$
&      2.614$\cdot10^{-2}$       ($\pm$0.034$\cdot10^{-2}$)&
2.613$\cdot10^{-       2}$      &       2.613$\cdot10^{-2}$&
2.611$\cdot10^{-2}$&                    2.603$\cdot10^{-2}$&
2.610$\cdot10^{-2}$  &  2.603$\cdot10^{-2}$\\  \hline \hline
$R$, $\delta R$ & 0.0045$^{**}$,  0.0018 & 0.0215 & 0.0085 &
0.0067 & 0.0064 & 0.0049 & 0.0078  \\  \hline  \end{tabular}
\end{center}

$^*$ values calculated using the shell model (see text)

$^{**}$  R-factor  evaluated  on the  basis of  experimental
uncertainties $\sigma _{\bf G}$

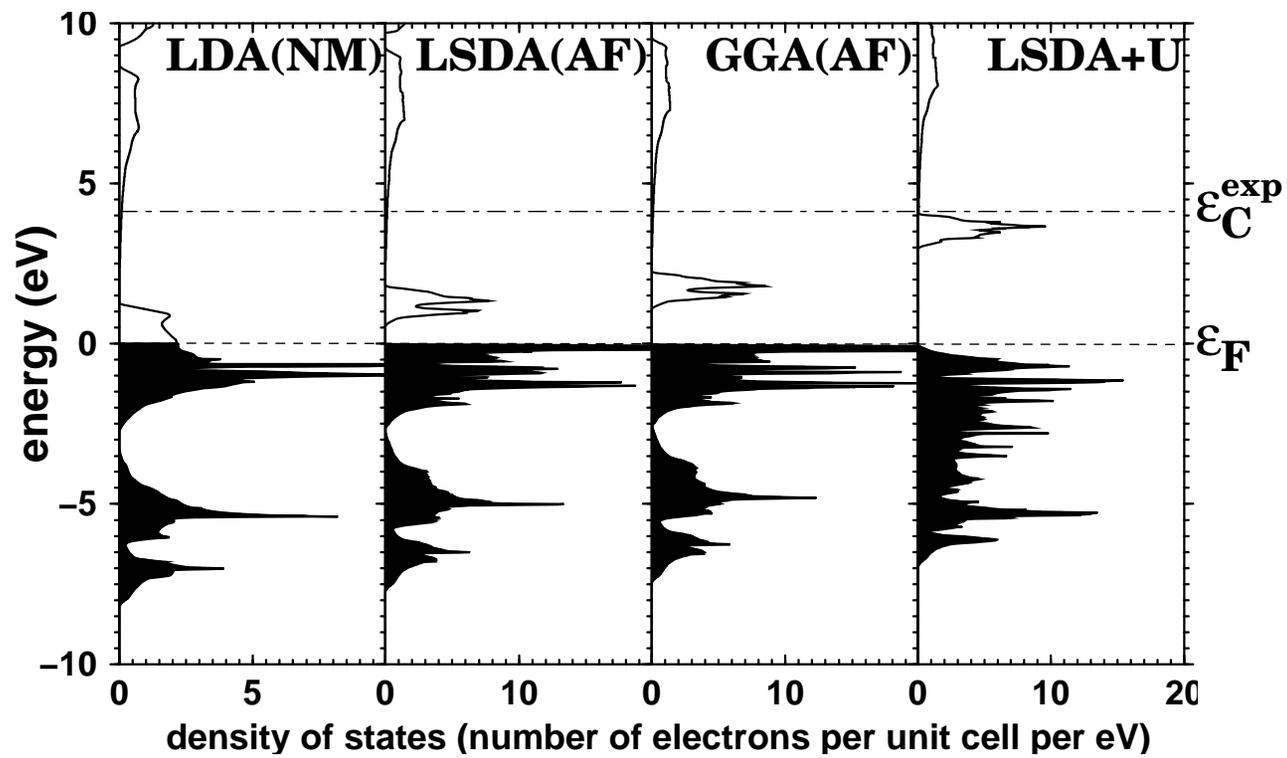
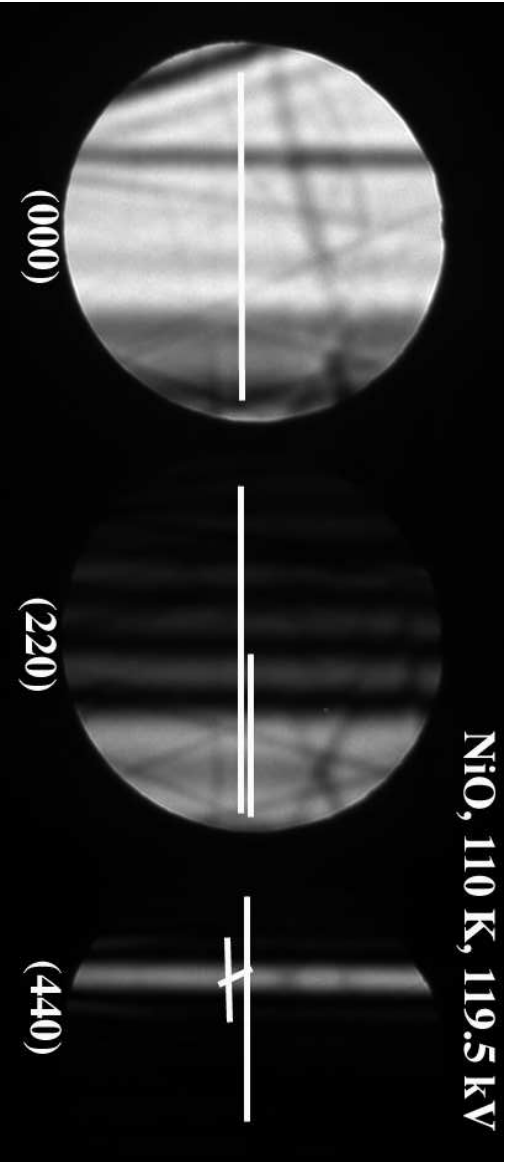
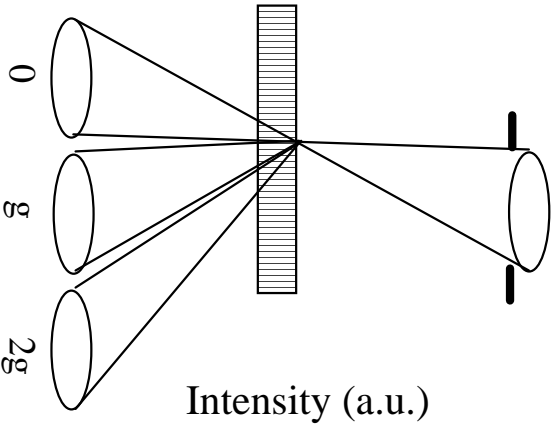
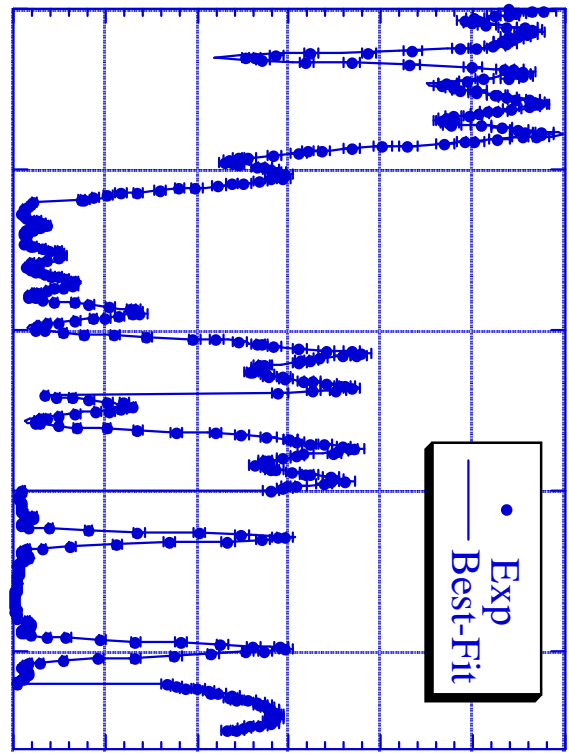
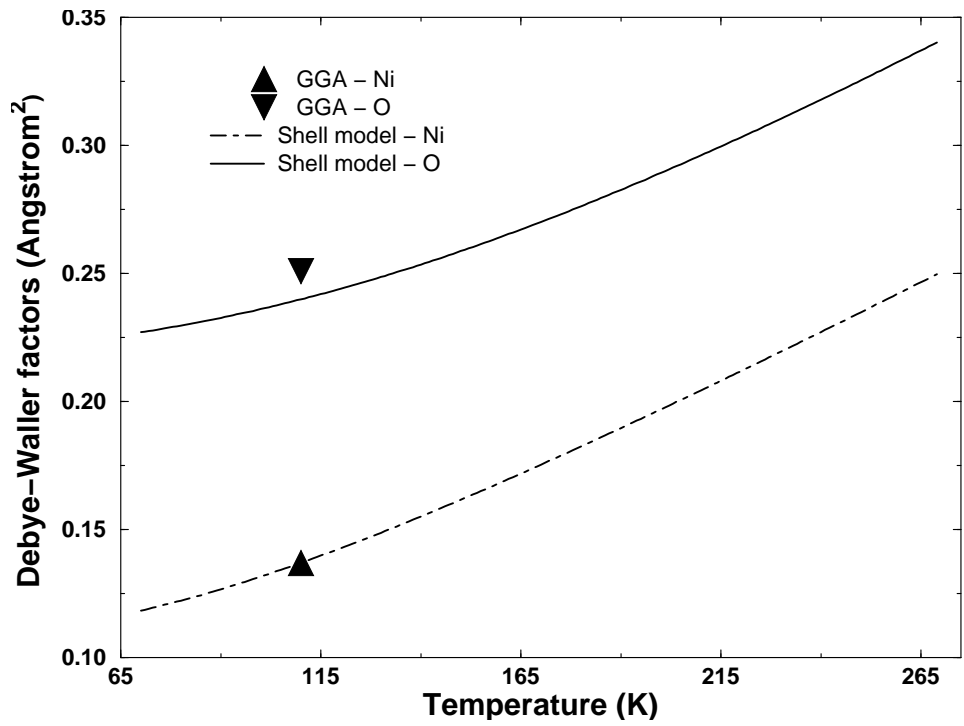
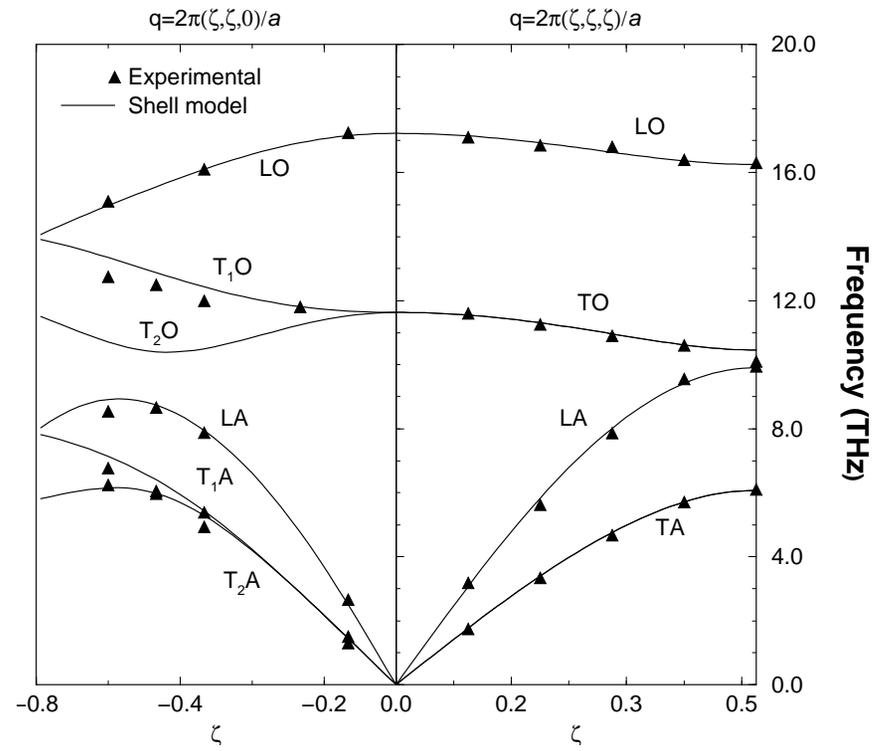
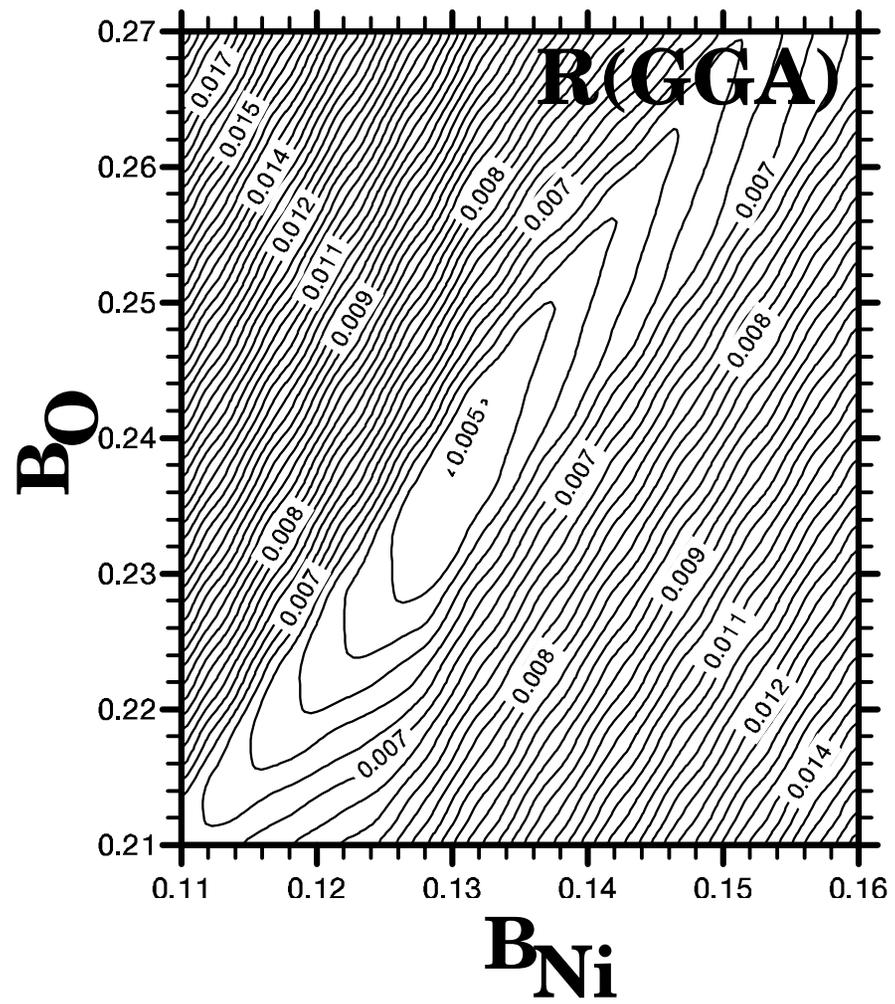
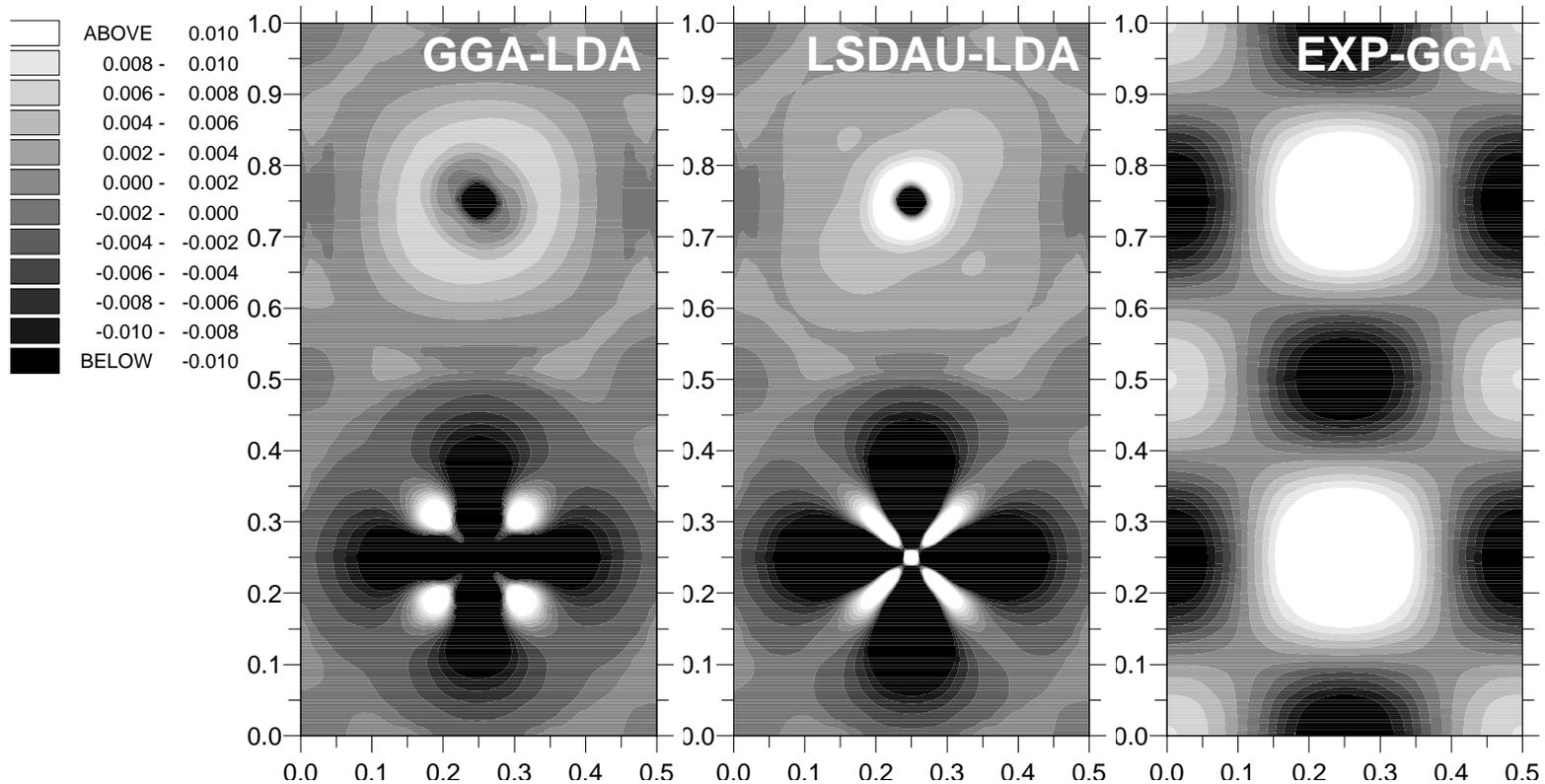
\end{document}